\documentclass[]{elsart}

\usepackage[dvips,final]{graphicx}
\usepackage{amsmath}
\usepackage{amssymb}
\usepackage[T1]{fontenc}
\usepackage{epsfig}

\begin{document}

\begin{frontmatter}

\title{A nuclear matter description based on quark structure of the nucleon and pion exchange}
\author{R. Huguet*, J.C. Caillon and J. Labarsouque}

\address{Centre d'Etudes Nucl\'{e}aires de Bordeaux-Gradignan, CNRS-IN2P3 \\ Universit\'{e} Bordeaux 1, Le Haut-Vigneau, 33170 Gradignan Cedex, France}

\begin{abstract}
We investigate the possibility to describe nuclear matter in an approach constrained by the prominent features of quantum chromodynamics. We mapped the in-medium nucleon self-energies of a point coupling relativistic mean-field model on self-energies obtained in effective theories of QCD. More precisely, the contributions to the nucleon self-energy have been separated into the short range part, driven principally by the quark structure of the nucleon described in a quark-diquark picture, and the long range part, dictated by pion dynamics and determined using in-medium chiral perturbation theory. A saturation point, although unrealistic, is obtained without any free parameter. A realistic description of nuclear matter saturation properties has been obtained with the inclusion of a small phenomenological correction term to the short range part of the self-energy.\\

\end{abstract}


\thanks[*]{Corresponding author \\ {\em e-mail address} : huguet@cenbg.in2p3.fr (R. Huguet)}
\end{frontmatter}

\section{Introduction} 

Effective field theories (EFT)  based on hadronic degrees of freedom are at the basis of many successful models of nuclear structure. In particular, quantum hadrodynamics (QHD) models offer a field theoretical framework consistent with the symmetries of quantum chromodynamics (QCD).  Effective Lagrangians, constructed with  meson-nucleon or point-coupling interactions, have been applied successfully in the relativistic mean field approximation (RMF) to describe nuclear matter and finite nuclei all over the periodic table \cite{serotwalecka,ring}. 

The nuclear matter saturation arises in QHD models via a subtle cancellation between large Lorentz scalar and four-vector nucleon self-energies in the medium. There is evidence from nuclear spin-orbit splitting in finite nuclei that these self-energies should be of the order of hundreds of MeV. A recent many body calculation \cite{plohl}, using realistic nucleon-nucleon (NN) potentials, found that these large self-energies is a model independent fact, enforced by the structure of NN interaction. 

However, in QHD models, the Lagrangian parameters are determined in a purely phenomenological way to reproduce nuclear matter and finite nuclei properties. It would be desirable to get a clear connection of the nucleon self-energies to the underlying theory of strong interactions, QCD. On the one hand, this can be achieved, for example, with finite density QCD sum rules \cite{cohen}.  In this approach, the nucleon self-energy modifications in the medium are related to the modifications of the scalar quark condensate and quark density. However, these self-energies alone do not allow for a realistic description of saturation properties when taken into account in an effective hadronic model \cite{aguirre}. On the other hand, important progress have been made in describing the nucleon self-energies by using quark models, like for example the Nambu-Jona-Lasinio model \cite{bentz}. This type of model is well suited for the study of short range effects in the mean-field approximation.

Concerning the long-range part of the NN interaction, it is known long before QCD that pion plays an important role. In the medium, the inclusion of pion exchange between nucleons in quark models, like the NJL one, leads to important technical difficulties and still seems out of scope. Another approach is based on the concept of effective field theories  \cite{weinberg1} in which one has to write the most general Lagrangian compatible with symmetries of QCD. At low energy, pions and nucleons are the relevant degrees of freedom and the dynamics is controlled by the broken chiral symmetry. This is the basis of chiral perturbation theory \cite{weinberg2} (ChPT), which has been successfully applied, for example, to construct NN potentials (see for examples \cite{ordo,machleidt,kaiser}). In this theory, the low-energy expansion of the Lagrangian is arranged in powers of derivatives and pion masses. The in-medium nucleon self-energies can in principle be determined with the in-medium ChPT as used first in \cite{kaiser2}. However, the determination of the contributions beyond the next-to-leading order in the medium is a really complicated task. Moreover, contact terms encoding the short distance dynamics arise at each order. 

In this work, we therefore investigate the possibility to obtain the nucleon self-energy in nuclear matter by using the complementary of these two approaches. The one and two pion exchange contributions are determined at next-to-leading order in the framework of in-medium ChPT. Instead of including the phenomenological contact terms, the short distance physics is described with a quark model of the nucleon. We use here a quark-diquark picture in a NJL model.
 
This nucleon self-energy is then implemented in a nucleonic density dependent point coupling model in order to explore the possibility to describe saturation properties of nuclear matter. A saturation point is obtained, although at too low density, for some values of the quark mass without any free parameter. Reasonable saturation properties are obtained by adding a small correction to the quark-diquark picture.

In section 2, we present the in-medium quark-diquark model of the nucleon in the NJL model. One and two pion exchange contributions to the nucleon self-energies are determined with in-medium ChPT in section 3. The nucleonic point coupling model, the self-energies mapping procedure are detailed in section 4, together with the results and discussion. We conclude in section 5. 

\section{In-medium self-energies of the nucleon in a Nambu-Jona-Lasinio quark-diquark picture} 

We derive in this section the in-medium modifications of the part of the nucleon self-energies associated to the internal quark structure of the nucleon. We use here a simple quark-diquark picture in NJL model. 

The Nambu-Jona-Lasinio \cite{njl}  model is a chirally invariant point coupling quark model, in which the dynamics is dictated by the dynamical breaking of chiral symmetry. The Lagrangian of two flavor ${\rm SU(2)}$ NJL model reads in a general form: 

\begin{equation}
\mathcal{L}_{\rm NJL} =\bar q\left[ i\FMslash{\partial}-m_{0}\right]q + \mathcal{L}^{\rm int},
\end{equation}

\noindent where $q$ is the flavour SU(2) quark field and $m_0$ the bare quark mass. $\mathcal{L}^{\rm int}$ is the interaction Lagrangian, based on chirally invariant four Fermi interactions. Any four fermion interaction can be Fierz symmetrised and rewritten identically as a chirally symmetric linear combinations $\sum_i g_i (\bar q \Gamma_i q)^2$ where $g_i$ are functions of the original couplings and $\Gamma_i$ are matrices in Dirac, flavor and color space. We consider the scalar, pseudoscalar and vector channels in order to reproduce the vacuum phenomenology of pion and omega mesons. The quark-quark interaction channels are similarly obtained by a decomposition of  $\mathcal{L}^{\rm int}$ using Fierz transformation. For our purpose, we need only the scalar diquark channel, which gives the leading contribution to the structure of the nucleon \cite{buck}.  We thus write the effective interaction Lagrangians in the $\bar qq$ and $qq$ channels respectively: 

\begin{eqnarray}
\mathcal{L}^{\rm int,eff}_{\bar qq} & = & g_{\pi}\left[ (\overline{q}q)^{2}+(\overline{q}i\gamma _{5}{\bf \tau }q)^{2}\right] -g_{\omega}(\overline{q}\gamma _{\mu }q)^{2}, \label{njlqantiq}    \\ 
\mathcal{L}^{\rm int,eff}_{qq} &= &  g_{sd}\left [ \bar q \beta^{A}\gamma^{5} C \tau_2 \bar q^{T} \right ] \left [q^{T} C^{-1} \tau_2 \beta^{A} \gamma_{5} q \right], \label{njlqq}
\end{eqnarray}

\noindent where $C=i\gamma_2 \gamma_0$, and $\beta^{A} = \sqrt{3/2}\lambda^A$  (A=3,5,7) are the color $\bar 3$ matrices.  The couplings $g_{\pi}$ and $g_{\omega}$, together with $m_0$, will be chosen to reproduce both the vacuum pion and vector masses and the pion decay constant. The scalar diquark coupling $g_{sd}$ is in principle related, via Fierz transformations, to the mesonic channel couplings. However, the form of this relationship depends on the choice of $\mathcal{L}^{\rm int}$ and is therefore not unique (see for example \cite {ishii}). As usual, we consequently have chosen the $g_{sd}$ value to reproduce the vacuum nucleon mass, independently from the other couplings.   

\subsection{Nucleon and mesons in vacuum}

The Dirac equation for a quark in mean-field approximation is given by:

\begin{equation}
\left[ i\FMslash{\partial}-m_{0}-2 g_{\omega}\gamma_{0}\left\langle \overline{q} \gamma^0 q \right\rangle +2g_{\pi}\left\langle \overline{q}
q\right\rangle \right] q=0,  \label{dir}
\end{equation}

\noindent which defines a dynamical constituent-quark mass:

\begin{equation}
m=m_{0}-2g_{\pi} \left\langle \overline{q}q\right\rangle, \label{gap}
\end{equation}

\noindent generated by a strong scalar interaction of the quark with the QCD vacuum. In the gap equation (Eq.(\ref{gap})), the quark condensate $\left\langle \overline{q}q\right\rangle $ can be written as:

\begin{equation}
\left\langle \overline{q}q\right\rangle =-\int \frac{d^{4}k}{\left( 2\pi
\right) ^{4}}\texttt{Tr}S(k),  \label{qqbs}
\end{equation}

where here Tr denotes traces over color, flavor and spin. In Eq.(\ref{qqbs}), $S(k)$ represents the quark propagator defined as:

\begin{equation}
S(k)=\frac{i}{\FMslash{k}-m+i\varepsilon }. \label{pro}
\end{equation}

\noindent The quark condensate is divergent due to the loop integral and requires an appropriate regularization procedure. As many authors\cite{b.m,bub}, we introduce a three-momentum cut-off $\Lambda $ which will have the least impact on medium parts of the regularized integrals, in particular at zero temperature\cite{bub}. In fact, since the model is non renormalizable, the cut-off $\Lambda$ is just an additional parameter. 

As usual, the mesons are obtained by solving the Bethe-Salpeter (BS) equation in the quark-antiquark channels.  Since this is the standard procedure, we only recall the principal steps. First, we define the quark-antiquark polarization operator in the $M=\pi ,\sigma ,\omega $ channel by:

\begin{equation}
\Pi _{M}(q^{2})=-i\int \frac{d^{4}k}{\left( 2\pi \right) ^{4}}\texttt{Tr}
\left[ \Gamma _{M}S(k+q/2)\Gamma _{M}S(k-q/2)\right] ,  \label{pola}
\end{equation}

where the vertex $\Gamma _{M}=i\gamma _{5}\tau ^{k}$, $i \mathit{1}$ and $i\gamma^{\mu }$ stand for respectively the pion, sigma and omega mesons. Note that, in the vector channel, the Lorentz structure of the polarization operator is $\Pi _{\omega }^{\mu \upsilon }(q^{2})=\left( -g^{\mu \nu }+\frac{q^{\mu}q^{\upsilon }}{q^{2}}\right) \Pi _{\omega }(q^{2})$. The meson masses, $m_{M}$, and meson-quark-quark coupling constants, $g_{Mqq}$, are then determined by the pole structure of the $T$-matrix, i.e. by the conditions:

\begin{equation}
1-K_{M}\Pi _{M}(q^{2}=m_{M}^{2})=0,  \label{pol}
\end{equation}

\begin{equation}
g_{Mqq}^{2}=\left[ \frac{d\Pi _{M}(q^{2})}{dq^{2}}\right]
_{q^{2}=m_{M}^{2}}^{-1},  \label{res}
\end{equation}

\noindent where $K_{M}=2g_{\pi}$, $2g_{\pi}$ and $2g_{\omega}$ respectively for the $\pi$, $\sigma$ and $\omega$ mesons.

The nucleon is described as a quark-diquark bound state by solving the Faddeev equation. The quark-quark bubble graph in scalar channel reads: 

\begin{equation} 
\Pi_{sd}(q) = 6i\int{\frac{d^4 k}{(2\pi)^4}\mathrm{Tr}_D \left [\gamma_5S(k)\gamma_5S(q-k) \right ]}, \label{polsd}
\end{equation} 

\noindent where ${\rm Tr_D}$ denotes a trace in Dirac space. The BS equation in the scalar diquark channel is then given by: 

\begin{equation}
 1 - 2 g_{sd}\Pi_{sd}(q) = 0.
\end{equation}

\noindent The relativistic Faddeev equation is reduced to an effective BS equation for a diquark and a quark interacting via quark exchange when restricting the qq channels to the scalar one. We take the static approximation to the Faddeev equation \cite{buck}, where the exchanged quark propagator is taken as $-i/m$. The static approximation has been found to reproduce reasonably well the exact Faddeev result \cite{ishii2}.  The Faddeev equation then reduces to a series of quark-diquark bubble graphs and the solution for the {\it t}-matrix in the color singlet channel is:  

\begin{equation} 
T(p) = \frac{3}{m}\frac{1}{1+\frac{3}{m}\Pi_N(p)},
\end{equation} 

\noindent with $\Pi_N(p)$ the quark-diquark ``polarization'': 

\begin{equation} 
\Pi_N(p) = i\int{\frac{d^4k}{(2\pi)^4}S(k)\tau(p-k)}, \label{poln}
\end{equation}

\noindent in which $\tau(q)$ is the diquark propagator 

\begin{equation} 
\tau(q) = \frac{4ig_{sd}}{1-2g_{sd}\Pi_{sd}(q)} \sim i\left[4g_{sd} + \frac{G_{sd}}{q^2-m_{sd}^2+i\epsilon}\right], \label{diprop}
\end{equation}

\noindent where the second equality results from the pole approximation of the propagator used in the numerical calculation. We have verified that this is a very good approximation of the exact diquark propagator in the range of momentum considered here. In Eq. (\ref{diprop}), $m_{sd}$ is the diquark mass, and $G_{sd}$ is defined as: 

\begin{equation} 
G_{\rm sd}^{2}=\left[ \frac{d\Pi _{\rm sd}(q)}{dq^{2}}\right]^{-1}_{q^{2}=m_{\rm sd}^{2}},  \label{diquarkcoupling}
\end{equation}

Note that for the determination of the polarizations (Eqs. (\ref{pola}), (\ref{polsd}) and (\ref{poln})), we have used the same regularization procedure as for the quark condensate.

\subsection{Nucleon self-energy modification in the medium} 

 As outlined by Ripka in \cite{ripka}, the in-medium modification of the quark condensate is roughly the same in quark or nuclear matter in NJL model, as long as the vacuum constituent quark mass is not too high ( $m \sim 400 {\rm MeV}$). In first approximation, we thus consider the in-medium modification of the quark-diquark state in quark matter.
The masses and couplings at finite density are denoted with star superscript (as for example $m^*$ for the constituent quark mass at finite density). When working at finite quark density, one has to modify the quark propagator Eq. (\ref{pro}):

\begin{itemize} 
\item by making the replacements $k_{\mu} \rightarrow k^*_{\mu}=k_{\mu}- 2g_{\omega} \left\langle \overline{q} \gamma_{\mu} q \right\rangle $ and $m \rightarrow m^*$
\item by adding to the quark propagator Eq. (\ref{pro}) the medium insertion part
\begin{equation}
-\pi \frac{\FMslash{k}^{*}+m^*}{E^*_{k}}\delta \left(k_{0}-E_{k}\right) \theta \left(k_{F}-\left| {\bf k}\right| \right), 
\end{equation} 
\noindent where $k_F$ is the quark Fermi momentum, and $E_k = E^*_k + 2g_{\omega}\rho$ with $\rho = \left \langle \bar q \gamma^0 q\right \rangle$ the quark density and $E^*_k= k_0^* = \sqrt{{\bf k^2}+m^{*2}}$.
\end{itemize}

\noindent The insertion of the medium part in each quark propagator generates finite integrals which can be performed almost analytically. The polarizations $\Pi_{j}$ ($j= M,sd,N$) in the Bethe-Salpeter and Faddeev equations thus depend explicitly on the quark Fermi momentum $k_F$. This explicit dependence in $k_F$ will not be indicated below.

 For a given Fermi momentum $k_F$, the product $ S(k)S(q-k)$ in Eq.(\ref{polsd}) for the diquark is replaced in the medium by $ S(k^*)S(q_{sd}-k^*)$, where $q^0_{sd}  = q^0 - 4g_{\omega}\rho$ and ${\bf q_{sd} = q }$. If  the equation $1 -2g_{sd}\Pi_{sd}(q^0)=0$ is satisfied, then the solution of  $1 -2g_{sd}\Pi_{sd}(q_{sd}^0)=0$  with $q_{sd} = (\sqrt{{\bf q}^2+m^{*2}_{sd}} = q^0-4 g_{\omega}\rho,{\bf q})$ is, in the rest frame of the diquark, $q^0 = m^*_{sd}+ 4g_{\omega}\rho$. 

Considering the nucleon ``polarization'' equation (\ref{poln}), $S(k)\tau(p-k)$ is replaced in the medium by $S(k^*)\tau(p_{d}-k^*)$ with $p_{d}^0=p^0-6g_{\omega}\rho$ and ${\bf p_d = p}$. If  the equation $ 1+ \frac{3}{m^*}\Pi_N(p) = 0$ is satisfied at $\FMslash{p} = M_N+\Sigma_S^0$, where $\Sigma_S^0$ is the scalar self-energy, then the solution of $ 1+ \frac{3}{m^*}\Pi_N(p_d) = 0$ is $\FMslash{p}_d = M_N+\Sigma_S^0$. We thus get $p^0 = \sqrt{{\bf p}^2+(M_N+\Sigma_S^0)^{2}} + 6g_{\omega}\rho$. The scalar self-energy $\Sigma_S^0$ is obtained by the solution of the quark-diquark pole condition, at rest (${\bf p=0}$), in quark matter. The vector self-energy is $\Sigma_V^0 = 6g_{\omega}\rho $. 

\noindent  Repeating this procedure for each value of $k_F$, we thus obtain $\Sigma_S^0(\rho_B)$ and $\Sigma_V^0(\rho_B)$ 
 
\subsection{Numerical results} 

For a given cut-off $\Lambda$, or equivalently a vacuum constituent quark mass $m$, the free parameters $m_{0}$, $g_{\pi}$ and $g_{\omega}$ are chosen to reproduce vacuum pion mass, pion decay constant and omega meson mass $ m_{\pi} = 135 {\rm MeV}$, $f_ {\pi} = 92.4 {\rm MeV}$, $ m_{\omega} = 782 {\rm MeV}$. The parameter $g_{sd}$ is finally taken to reproduce the free nucleon mass $M_N = 939 {\rm MeV}$. As usual, one has to consider several values for the vacuum constituent quark mass $m$.

\begin{figure}[htb]
	{\centering
		\epsfig{file=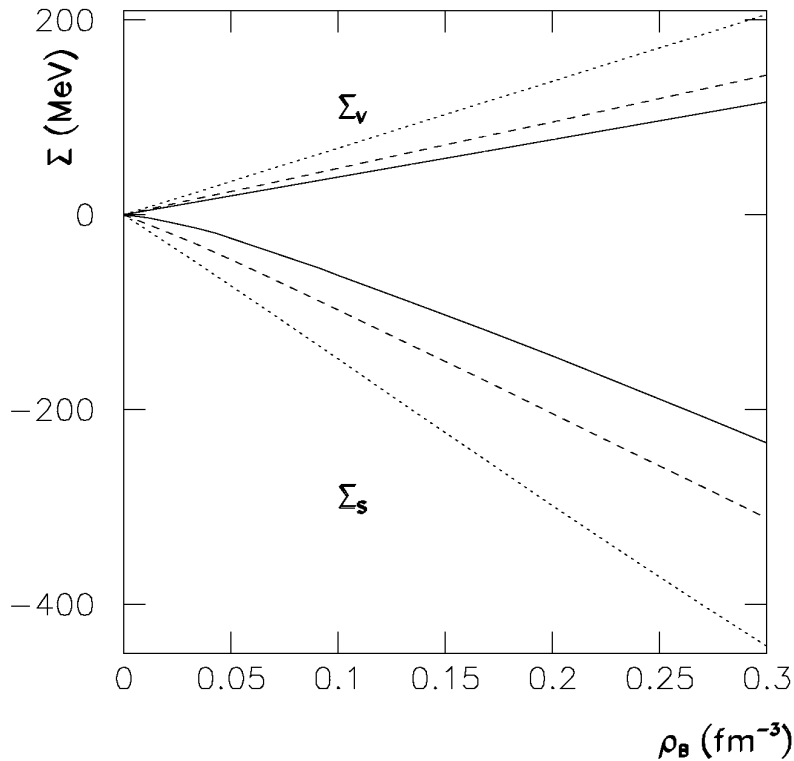,scale=0.95}
	\caption{ Part of the scalar and vector nucleon self-energies from quark-diquark model, for three values of the vacuum constituent quark mass: $m=370 {\rm MeV}$ (solid line), $m=400 {\rm MeV}$ (dashed line) and $m=430 {\rm MeV}$ (dotted line). }
	\label{quarkdiquarkself}}
\end{figure}

The results for the in-medium nucleon self-energies are reported on Fig. \ref{quarkdiquarkself} for three values of the constituent quark mass between $ m= 370 {\rm MeV}$ and $m=430 {\rm MeV}$. For $m<370 {\rm MeV}$, the nucleon is not bound and for $m>440 {\rm MeV}$ the scalar self-energy of the nucleon is too much attractive. The parameters of the models and the values of the nucleon self-energies at a baryonic density $\rho_0 \sim 0.17 {\rm fm}^{-3}$ are given in Tab. \ref{qdtab}.

\begin{table}[b]
\begin{center}
\begin{tabular}{cccccccc}
\hline\hline
$m {\rm (MeV)}$ & $\Lambda {\rm (MeV)}$ & $m_0 {\rm (MeV)}$ & $g_{\pi}\Lambda^2$ & $g_{\omega}\Lambda^2$ & $g_{\rm sd}/g_{\pi}$ & $\Sigma_S^{0} {\rm (MeV)}$ & $\Sigma_V^0  {\rm (MeV)}$ \\ 
\hline
370 & 605 &  5.9 & 2.3 & 2.0 & 0.735 & -260 & 130 \\ 
400 & 592 & 6.0 & 2.4 & 2.4 & 0.768 & -347 & 160 \\ 
430 & 583 & 6.0 & 2.4 & 3.3 & 0.789 & -490 & 230 \\ 
\hline\hline
\end{tabular}
\caption{NJL model parameters for three values of the constituent quark mass. The nucleon self-energies in the last two columns are given at the baryonic density $\rho_0 = 0.17 {\rm fm^{-3}}$. \label{qdtab}}  
\end{center} 
\end{table} 

Some comments are in order, even if these results should not be interpreted alone. First, the scalar self-energy is attractive and the vector one is repulsive as expected. However, one interesting point is that there is a strong asymmetry between the scalar and vector part of the self-energy, with a ratio $\Sigma_S^0/\Sigma_V^0 \sim -2$. 

Second, the vector self-energy is exactly linear in the baryonic density. As can be seen in Fig. \ref{quarkdiquarkself} and verified numerically, the scalar self-energy is also linear in the density to a good approximation.

Finally, the self-energies highly depend on the vacuum quark mass value. There is approximately a factor $1/2$ from $m=370 {\rm MeV}$ to $m=430 {\rm MeV}$ for the part of the scalar self-energy coming from the quark structure of the nucleon. However, we will show later that the impact of the quark mass value on the nuclear matter saturation properties is not as crucial as it could be expected from these results. 

\section{In-medium nucleon self-energies from pion exchange contributions} 

We derive in this section the in-medium modifications of the nucleon self-energies coming from the long-range pion exchange determined in the framework of chiral perturbation theory (ChPT).

\subsection{Elements of chiral perturbation theory}

ChPT is an EFT of pions and nucleons, constrained by QCD chiral symmetry. Numerous reviews on ChPT applied to NN interaction (see for example \cite{machleidt2}) or nuclear matter (see for example \cite{kaiser2,opticpot}) are available. We will recall here only the principal features needed for our discussion. 

Since we need only pion exchange contributions, we consider only the $\pi N$ interaction Lagrangian. Since the interactions of the Goldstone bosons (pions) must vanish at zero momentum transfer and in the chiral limit ($m_{\pi} \rightarrow 0$), the low energy expansion of the ChPT Lagrangian is arranged in powers of derivatives and pion masses:

\begin{equation}
\mathcal{L}_{\pi N} = \mathcal{L}_{\pi N}^{(1)} + \mathcal{L}_{\pi N}^{(2)} + \ldots
\end{equation}

\noindent The lowest order $\pi N$ Lagrangian, with one derivative or pion mass insertion, can be written: 

\begin{equation} 
\mathcal{L}_{\pi N}^{(1)} = \bar \psi (i\FMslash{D} - M_N + \frac{g_A}{2}\gamma^{\mu} \gamma_5 u_{\mu})\psi,
\end{equation}

\noindent where $\psi$ is the nucleon Dirac spinor, and: 

\begin{eqnarray} 
D_{\mu} &=& \partial_{\mu} + \Gamma_{\mu}, \\ 
\Gamma_{\mu} & =& \frac{1}{2}\left ( \xi^{\dagger}\partial_{\mu}\xi + \xi\partial_{\mu}\xi^{\dagger}\right )=\frac{i}{4f_{\pi}^2}\boldsymbol{\tau}.(\boldsymbol{\pi} \times\partial_{\mu} \boldsymbol{\pi})\ldots \\ 
u_{\mu} & = & i\left (\xi^{\dagger}\partial_{\mu}\xi -\xi \partial_{\mu}\xi^{\dagger}\right )=-\frac{1}{f_{\pi}}\boldsymbol{\tau}.\partial_{\mu}\boldsymbol{\pi} + \ldots \\
U &=& \xi^2 = 1 + \frac{i}{f_{\pi}}\boldsymbol{\tau}.\boldsymbol{\pi} - \frac{1}{2f_{\pi}^2}\boldsymbol{\pi}^2 + \ldots 
\end{eqnarray}

\noindent with $U=\xi^2$ the $SU(2)$ matrix collecting pion fields. 

The contributions in ChPT are usually analyzed in terms of the powers of small momentum over the large scale $(Q/\Lambda_{\chi})^{\nu}$, where $Q$ stands for the characteristic momentum of the problem under consideration (nucleon or pion momentum, pion mass, \ldots) and $\Lambda_{\chi}\sim 1 \ {\rm GeV}$ is the chiral symmetry breaking scale. The chiral counting rule consists in determining at which power $\nu$ each Feynman diagram contributes. 

For the self-energy diagrams we consider, the chiral counting rule gives rise to the following hierarchy of contributions: 

\begin{itemize} 
\item $\nu=3$ is the lowest order (denoted order $\mathcal{O}(Q^3)$), given by the one pion exchange diagram (one loop) with vertex insertion from $\mathcal{L}_{\pi N}^{(1)}$; 
\item $\nu=4$ order (denoted order $\mathcal{O}(Q^4)$) receives contribution from the iterated one pion exchange diagram (two loops) with vertex insertion from $\mathcal{L}_{\pi N}^{(1)}$;
\item $\nu=5$ order (denoted order $\mathcal{O}(Q^5)$) gets contributions also from the iterated one pion exchange diagram, and irreducible two pion exchange diagrams, with vertex insertion from $\mathcal{L}_{\pi N}^{(1)}$;
\item at $\nu=6$ order, two pion exchange diagrams with one vertex insertion from the second order Lagrangian $\mathcal{L}_{\pi N}^{(2)}$ (with two derivatives or pion mass insertions) and the other vertices from $\mathcal{L}_{\pi N}^{(1)} $, contribute;
\item \ldots
\end{itemize}

\noindent Considering the technical difficulties of in-medium calculations, we will restrict ourselves to one pion (Fig. \ref{1pe}) and iterated one pion exchange diagrams (Figs. \ref{ipehartree} and \ref{ipefock}).  We thus will determine contributions at order $\mathcal{O}(Q^5)$.  Irreducible two pion exchange diagrams contribute for only a few MeV \cite{opticpot}, and we will thus neglect these contributions in this first approach.

Finally, the Heavy Baryon formulation of ChPT \cite{hbchpt} is often used for NN interaction. It consists in using a $1/M_N$ expanded Lagrangian instead of $\mathcal{L}_{\pi N}^{(1)}$. However, we will use the equivalent method of the Munich group \cite{kaiser}. One starts with the relativistic Lagrangian $\mathcal{L}_{\pi N}^{(1)}$ and writes down relativistic amplitudes. Relativistic vertices and propagators are then expanded in $1/M_N$. This method leads to the same results obtained when starting with a $1/M_N$ expanded Lagrangian, but is more efficient in dealing with the calculations. 

\subsection{In-medium self-energy diagrams} 

For the in-medium self-energy diagrams, the relevant small scale of the problem is the Fermi momentum $k_F$. Thus, the chiral hierarchy exposed above suggests that we determine contributions at order $\mathcal{O}(k_F^5)$.

Denoting $k = (k^0=\sqrt{{\bf k}^2+M_N^2},{\bf k})$ the four vector of the on-shell in-going nucleon, the self-energy for each of the diagrams considered can be written as: 

\begin{equation} 
\Sigma(k) = \Sigma_S(k) + \gamma_0\Sigma_V(k) - {\boldsymbol \gamma }. {\bf k} \Sigma_{v}(k)
\end{equation}

 As argued in \cite{horowitz,terhaar}, it is known from relativistic Hartree-Fock and Dirac-Bruckner calculations that the real part of the spatial component $|{\bf k}|{\Sigma_v}$ is much smaller than $\Sigma_S$, $\Sigma_V$. Moreover, in the case of pion exchange with pseudo-vector coupling considered here,  $|{\bf k}\Sigma_v|/\Sigma_S $ is of order $ |{\bf k}|/M_N$, with $|{\bf k}|<k_F$, for all the diagrams considered. We will thus neglect the spatial component $\Sigma_v$ of the self-energy. 

We determine the in-medium contributions arising from the self-energy diagrams using the procedure described in \cite{kaiser2,opticpot}. We recall here the principal features of this method. The diagrammatic calculation in nuclear matter involves the in-medium nucleon propagator:

\begin{eqnarray} 
S_N(k) &= & (\FMslash{k} + M_N) \left \{ \frac{i}{k^2-M_N^2+i\epsilon} -2\pi\delta(k^2-M_N^2)\theta(k_0)\theta(k_F -|{\bf k}|)\right \}, \nonumber \\ 
 & = & S_N^{\rm vac}(k) + S_N^{\rm med}(k),
\end{eqnarray} 

\noindent the pion propagator: 
\begin{equation} 
G_{\pi}(q) = \frac{i}{q^2-m_{\pi}^2+i\epsilon},
\end{equation} 
\noindent and the $\pi-NN$ vertex from $\mathcal{L}_{\pi N}^{(1)}$: 

\begin{equation} 
\frac{g_{A}}{2f_{\pi}}\FMslash{q}\gamma_5\vec\tau.
\end{equation}

 \noindent  As usually done \cite{kaiser2,lutz}, the pion exchange self-energy diagrams are considered as perturbations to the vacuum. Consequently, the masses and couplings involved should be taken at their vacuum physical value.

\noindent The nucleon propagator splits additively into the vacuum $S_N^{\rm vac}$ and medium $S_N^{\rm med}$ insertion parts. The calculation is organized according to the number of medium insertions.
Self-energy diagrams with no medium insertion part contribute to the physical mass of the nucleon $M_N$. Therefore, we consider only diagrams with one or more medium insertions. 

\subsubsection{One pion exchange diagram} 

\begin{figure}[htb]
	{\centering
		\epsfig{file=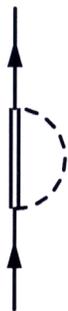,scale=1.0}
	\caption{One pion exchange Fock diagram.}
	\label{1pe}}
\end{figure}

The one pion exchange Hartree diagram is trivially zero. The one pion exchange Fock diagram, with one medium insertion on the internal nucleon line is shown on Fig. \ref{1pe}. The double line on diagrams stands for a medium insertion part of the nucleon propagator, the simple line for a vacuum part and the dashed line stands for the pion propagator.  

\noindent One has to write the complete amplitude of this diagram. After performing the integration on the time component of internal four vector, one is left with a three dimensional integral. The integrand is expanded in $1/M_N$, keeping the two first orders. The remaining integrals can then be achieved. The resulting self-energies are functions of both $|{\bf k}|$ and $k_F$.  As usual \cite{finelli}, we will take its value for a nucleon on top of the Fermi sea with $|{\bf k}| =k_F$. We thus will quote here only the expressions of the self-energy contributions for $|{\bf k}|=k_F$. The contribution to the nucleon self-energy of the one pion exchange Fock diagram (denoted $\Sigma_S^1$, $\Sigma_V^1$) is given by: 

\begin{equation} 
\Sigma_S^{1}(v) =3\left(\frac{g_A}{4\pi f_{\pi}}\right)^2m_{\pi}^3\left \{ \Sigma_S^{1,1} + \frac{m_{\pi}^2}{M_N^2}\Sigma_S^{1, 2}\right \}, 
\end{equation} 

\noindent with:

\begin{equation} 
\Sigma_S^{1,1}(v)  = \frac{1}{4v}\ln\left(\frac{1}{4v^2+1}\right) +\frac{2}{3}v^3 -v+ \arctan(2v), 
\end{equation} 

\noindent and: 
\begin{equation}
\Sigma_S^{1,2}(v)=  \frac{-6v^2+1}{16v}\ln\left(\frac{1}{4v^2+1}\right) -\frac{v(v^2-1)}{2}\arctan(2v) -\frac{v^5}{5}-\frac{v^3}{6} -\frac{3v}{4}.
\end{equation}

\noindent where $v=k_F/m_{\pi}$. The convergence of the $1/M_N$ expansion can be verified at this stage. The $m_{\pi}^2/M_N^2$ contribution of $\Sigma_S^{1}$  indeed represents 5\% of the first order contribution, for a value of $k_F \approx 250 {\rm fm}^{-3}$ close to the saturation one.

\noindent Finally, $\Sigma_V^1$ is given by: 
\begin{equation} 
\Sigma_V^{1}(v) =3\left(\frac{g_A}{4\pi f_{\pi}}\right)^2m_{\pi}^3\left \{ \Sigma_S^{1}(v) +\frac{v^2m_{\pi}^2}{M_N^2}\Sigma_S^{1,1}(v)\right \}.
\end{equation} 

\subsubsection{Iterated one pion exchange diagram} 

\begin{figure}[htb]
	{\centering
		\epsfig{file=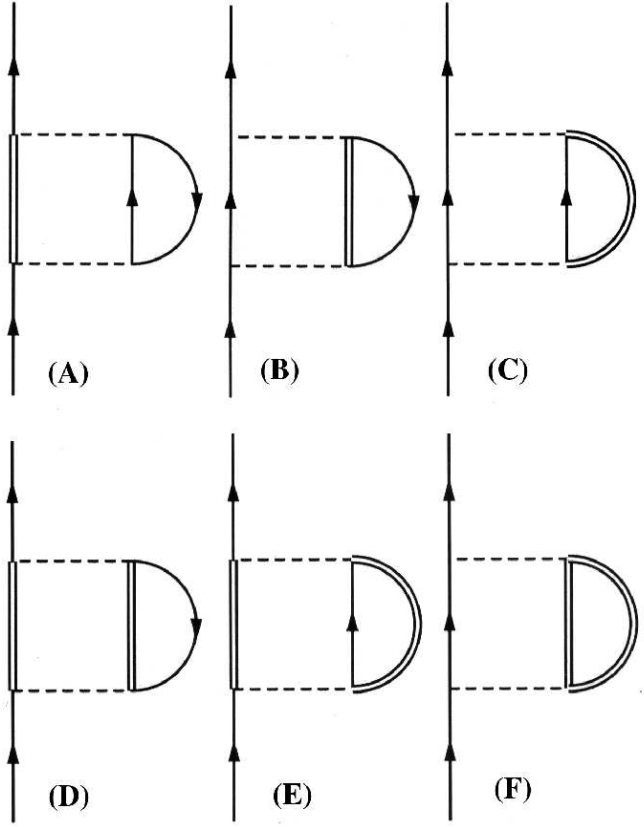,scale=1.5}
	\caption{Iterated one pion exchange Hartree diagrams with one or two medium insertions.}
	\label{ipehartree}}
\end{figure}

 The iterated one pion exchange Hartree diagrams with one or two medium insertions are represented on Fig. \ref{ipehartree}. The diagram with three medium insertions brings a purely imaginary part to the amplitude. The diagram { \sc A} contributes to the physical pion mass. The diagram {\sc B} is of order $\mathcal{O}(1/M_N^5)$ and is sub-dominant. Finally, there are four remaining one pion iterated exchange Hartree diagrams which contribute, namely {\sc C,D,E,F}.  

We focus here on the {\sc C} diagram with one medium insertion. First, the internal part of this diagram is the so called {\it planar box graph} which needs to be treated carefully. This is done in detail in \cite{kaiser}. The amplitude involves four poles in each half complex plane. The procedure consists in performing the internal $q^0$ integral by Cauchy's theorem, and then expanding the remaining integrands at first order in $1/M_N$ for each pole. The dominant contribution arises from one of the nucleon poles. Using the tools of dimensional regularization procedure, it is found that this diagram is not divergent and does not need any regularization as outlined in \cite{kaiser}.

 The introduction of a linear cut-off in the iterated one pion exchange contribution, as done in \cite{kaiser2,opticpot}, can be considered as an effective parameterization of contact terms. Alternatively, Lutz {\it et al.} choose to introduce explicitly a NN contact term in the Lagrangian in \cite{lutz}. Either the cut-off or NN contact term is a parameterization of the short range physics. One of the goals of this work is to try to get a description of this short range physics more directly from the quark structure of the nucleon, and thus, to avoid double counting, we have not introduced such contact terms or cut-off in the pionic part of the self-energies. The short range physics will be explicitly taken into account in our approach with the quark-diquark description of the nucleon. It is thus clear that the interpretation of our results should be made only when both the quark-diquark and pion exchange contributions to the self-energy are taken into account. 

The contributions to the nucleon scalar self-energy of the one pion iterated exchange Hartree diagrams C (denoted $\Sigma_S^2$) and (D+E+F) (denoted $\Sigma_S^3$) can finally be written in the form, for $|{\bf k}| =k_F$: 

\begin{eqnarray}
\Sigma^{2}_S(v) & =& -6\pi\left(\frac{g_Am_{\pi}}{4\pi f_{\pi}}\right )^4M_N  \nonumber \\ 
 & & \left \{  \frac{11}{6}v +\frac{7}{24v}\ln\left [1+4v^2\right ] -\left[\frac{3}{2}+\frac{4}{3}v^2\right]\arctan\left(2v\right)\right \},
\end{eqnarray}

\begin{eqnarray}
\Sigma_S^{3}(v) & = & 6\left(\frac{g_Am_{\pi}}{4\pi f_{\pi}}\right)^4M_N \left \{  \right . \nonumber \\
& &  v^2\int_{0}^1\frac{dw}{2}\left [ 8v^2w^2\frac{1+2v^2w^2}{1+4v^2w^2}-2 \ln\left (1+4v^2w^2\right)\right]  \nonumber \\ 
 & & \left (w+ \frac{1}{2}\left(1-w^2\right )\ln\left [\frac{w+1}{w-1}\right ]\right ) \nonumber \\  
 &  & + \int_0^1dw\int_{-vw}^{vw} d\xi  \left \{ \frac{(\xi+vw)^5 }{((\xi + vw)^2+1)^2}\right \} \nonumber \\ 
& &  \left (\xi v+ \frac{1}{2}\left( v^2-\xi^2\right )\ln\left [\frac{\xi+v}{\xi-v}\right ]\right ) \nonumber \\ 
 & & + \left .\int_{-1}^1dw\int_0^{v} \frac{\xi^2}{v}d\xi \left [ \frac{2\sigma^2+\sigma^4}{2(1+\sigma^2)}-\ln\left (1+\sigma^2\right) \right ]  \ln\left |\frac{\xi w+v}{\xi w-v}\right| \right \},
\end{eqnarray}

\noindent with  $\sigma = \xi w + \sqrt{v^2 + \xi^2(w^2-1)}$ and $v$ as defined earlier. The contribution to the vector self-energy of the nucleon is strictly equal to the scalar one.

\begin{figure}[htb]
	{\centering
		\epsfig{file=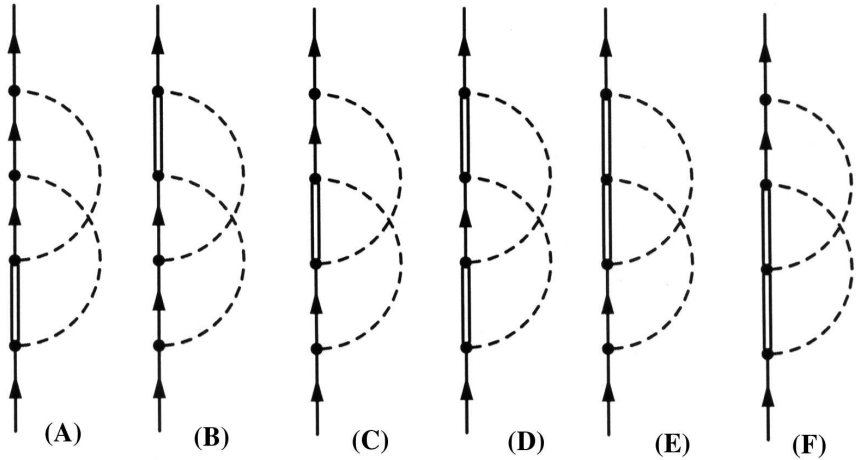,scale=1.0}
	\caption{Iterated one pion exchange Fock diagrams with one or two medium insertions.}
	\label{ipefock}}
\end{figure}

The iterated one pion exchange Fock diagrams with one or two medium insertions are represented on Fig. \ref{ipefock}. The diagram with three medium insertions brings a purely imaginary part to the amplitude. The diagrams { \sc A, B} contribute to the physical masses and couplings. Finally, there are four remaining one pion iterated exchange Fock diagrams which contribute, namely {\sc C,D,E,F}.   

\noindent The contributions to the scalar part of the nucleon self-energy from the iterated one pion exchange Fock diagrams C (denoted $\Sigma_S^4$) and (D+E+F) (denoted $\Sigma_S^5$) can be written as, for $|{\bf k}| = k_F$: 

\begin{eqnarray}
\Sigma^{4}_S(v) & = &  2\pi \left (\frac{g_Am_{\pi}}{4\pi f_{\pi}}\right )^4 M_N\left \{ v^3 + \int_0^{v} \xi(1-\frac{\xi}{v}) d\xi \frac{3}{1+2\xi^2} \right. \nonumber\\ 
& & \left. \left [ (1+8\xi^2+8\xi^4)\arctan \xi - (1+4\xi^2)\arctan(2\xi) \right ] \right \},
 \end{eqnarray}

\begin{eqnarray} 
  \Sigma^{5}_S(v) & = &  \frac{3}{2}\left (\frac{g_Am_{\pi}}{4\pi f_{\pi}}\right )^4M_N\left\{ \frac{G^2(v)}{8v^2} \right. \nonumber \\ 
 & & \left . + \int_0^v d\xi G(\xi) \left [1+\frac{\xi^2-v^2-1}{4v\xi}\ln\frac{1+(v+\xi)^2}{1+(v-\xi)^2}  \right ] \right .  \nonumber \\ 
  & & + \frac{1}{4}\int_{0}^1dw \int_{0}^1d\xi \frac{w}{z}\left (4v^2w^2-\ln(1+4v^2w^2) \right )\nonumber \\ 
 & & \left(-4v^2z^2+\ln(1+4v^2z^2)\right) \nonumber \\ 
& & + \int_{-1}^1dw\int_0^v d\xi\frac{\xi^2}{v}[\ln(1+\sigma^2)-\sigma^2]\left ( \ln \frac{|v+\xi w |}{|v-\xi w|}\right . \nonumber \\
 & &  \left . \left. + \frac{1}{R}\ln\frac{[vR+(v^2-\xi^2-1)w\xi]^2}{[1+(v+\xi)^2][1+(v-\xi)^2][v^2-\xi^2w^2]}  \right )  \right \},
\end{eqnarray} 

\noindent with $\sigma=\xi w + \sqrt{v^2 + \xi^2(w^2-1)}$ as defined earlier, and: 

\begin{eqnarray} 
G(x) &=& v(1+v^2+x^2) \nonumber \\ 
 & & - \frac{1}{4x}[1+(v+x)^2][1+(v-x)^2]\ln\frac{1+(v+x)^2}{1+(v-x)^2}, \\ 
R & =& \sqrt{(1+v^2-\xi^2)^2+4\xi^2(1-v^2)}, \\ 
z &= & \sqrt{(\xi^2-1)w^2+1}.
\end{eqnarray}

\noindent The vector part of the nucleon self-energy is again equal to the scalar part. 


\subsection{Numerical results} 

The physical input parameters are $ M_N=939 {\rm MeV}$, $m_{\pi} = 135 {\rm MeV}$, $f_{\pi} = 92.4 {\rm MeV}$ and $g_A = 1.3$. The self-energy coming from pion exchange is completely parameter free. Even if the expressions obtained above for the self-energies contain all powers of the parameter $k_F/m_{\pi}$, the chiral counting rule indicates that the diagrams taken into account contribute mainly at order $\mathcal{O}(k_F^3)$, $\mathcal{O}(k_F^4)$ and $\mathcal{O}(k_F^5)$. It is thus natural to take the scalar and vector self-energies as polynomial fits up to $k_F^5$: 

\begin{equation} 
\Sigma_{i}^{\pi}(k_F) = \left[c^{i}_0+ c^{i}_1\frac{\rho_B^{1/3}}{M_N} + c^{i}_2\frac{\rho_B^{2/3}}{M_N^2}\right ]\frac{\rho_B}{M_N^2}, \label{cij}
\end{equation}

\noindent where the $c_{j}^{i}$ are density independent parameters in the $i=S,V$ scalar or vector channels, and $\rho_B = \frac{2}{3\pi^2}k_F^3$ is the baryonic density. These parameters are reported in Tab. \ref{chpttab}. This numerical fit perfectly reproduces the exact self-energies. 

\begin{table}[htb]
\begin{center}
\begin{tabular}{ccc}
\hline\hline
 & S & V \\ 
\hline
$c_0$ & 40.5 & 40.5 \\ 
$c_1$ &  469.9& 468.8\\ 
$c_2$ & 252.9 & 283.3 \\ 
\hline\hline
\end{tabular}
\caption{Coefficients of the self-energies Eq. (\ref{cij}) for the pionic contribution in scalar (S) and vector channels (V). \label{chpttab}}  
\end{center} 
\end{table} 

\begin{figure}[htb]
	{\centering
		\epsfig{file=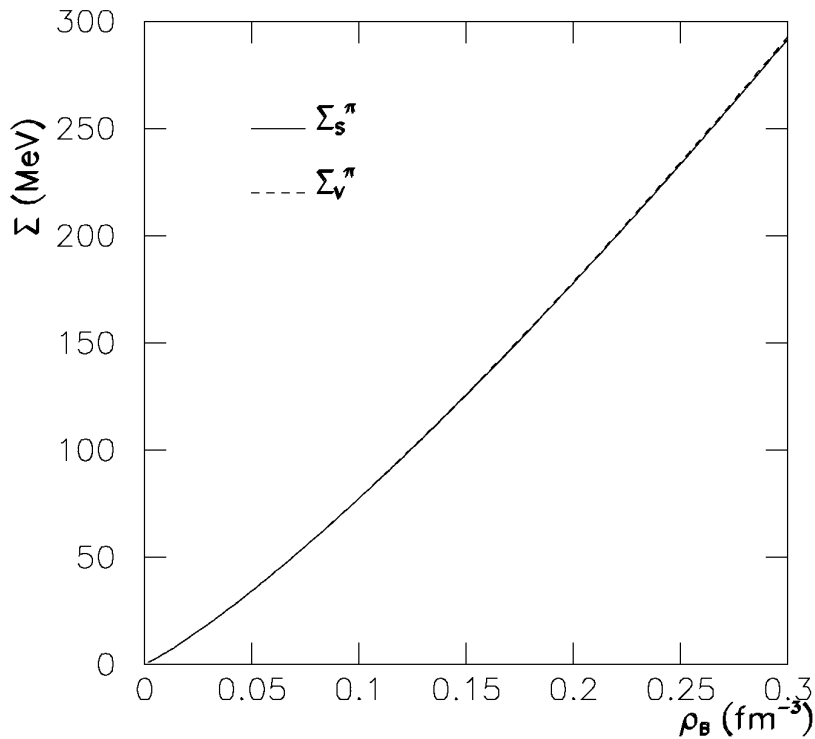,scale=0.95}
	\caption{Scalar and vector nucleon self-energies from ChPT at fifth order in $k_F$ as functions of baryonic density. }
	\label{chptself}}
\end{figure}

\noindent The self-energies are shown on Fig. \ref{chptself} as functions of the baryonic density.

\noindent Some comments are in order. First, the scalar and vector self-energies have very close values. This is not surprising since the iterated one pion diagrams are taken at the non relativistic order $\mathcal{O}(1/M_N)$, which implies that $\Sigma_S^{\pi} = \Sigma_V^{\pi}$. The small difference between the self-energies arises from the first relativistic $\mathcal{O}(1/M_N^2)$ correction taken in the one pion Fock diagram. 

Second, the part of the self-energies coming from pion exchanges is approximately $\Sigma_{S,V}(\rho_0 = 0.17{\rm fm^{-3}}) \sim 130 {\rm MeV}$, which gives a repulsive single particle potential. Let us recall that, at this point, these results include only the pionic part and that the short range physics is missing.

Finally, higher order contributions in ChPT could be important. These contributions would arise first as order $\mathcal{O}(k_F^6)$ contributions.

\section{Nuclear matter} 

We need a framework to describe nuclear matter in which the self-energies determined from nucleon structure in the quark-diquark model and pion exchange contributions will dictate the dynamics. It should be covariant and flexible enough to impose the dynamics as obtained in the quark-diquark and pion exchange calculations. 

\subsection{Density dependent point coupling model} 

A point-coupling nucleonic Lagrangian with density dependent couplings offers a framework compatible with the requirements described above. Such a framework, with a mapping procedure of the self-energies, has been already used in \cite{finelli} by Finelli et al. We are here interested only in the description of symmetric and infinite nuclear matter in the relativistic mean-field (RMF) approximation and the Lagrangian reads: 

\begin{eqnarray} 
\mathcal{L} & = & \bar \psi (i \gamma_{\mu}\partial^{\mu}  -M_N)\psi  \nonumber\\ 
 &  & -\frac{1}{2}G_S(\hat\rho_B) (\bar\psi \psi)^2 - \frac{1}{2}G_V(\hat\rho_B)(\bar\psi \gamma_{\mu} \psi)^2, 
\end{eqnarray}

\noindent where $\psi$ is the nucleon field, $\hat\rho_B$ is the baryonic density operator with $ \hat\rho_B u^{\mu} = \bar\psi\gamma^{\mu}\psi$ and $u^{\mu}=(1-v^2)^{-1/2}(1,{\bf v})$ is the quadri-velocity of the nuclear fluid. The coupling strengths $G_S$ and $G_V$ are supposed to be functional of the baryonic density. These functions will be determined by mapping the nucleon self-energies on those obtained from the quark-diquark and pion exchange calculations.  This Lagrangian has to be used in mean-field approximation, with fluctuations beyond the mean-field encoded in the density dependences. The equation of motion for the nucleon reads: 

\begin{equation}
\left [ \gamma_{\mu}(i\partial^{\mu} - V^{\mu}) -(M_N+S) \right ] = 0, 
\end{equation}

\noindent with the scalar and vector self-energies of the nucleon $S$ and $V^{\mu}$ given by 

\begin{eqnarray} 
S & = & G_S(\hat\rho_B)(\bar\psi\psi), \\ 
V^{\mu} &= &  G_V(\hat\rho_B)(\bar\psi\gamma^{\mu}\psi) + \Sigma^{\mu}_r, \\ 
\Sigma^{\mu}_r & =& \frac{u^{\mu}}{2}\left ( \frac{\partial G_S}{\partial\hat\rho_B}(\bar\psi\psi)^2+ \frac{\partial G_V}{\partial\hat\rho_B}(\bar\psi\gamma^{\mu}\psi)^2 \right). 
\end{eqnarray}
  
\noindent In the RMF approximation for infinite symmetric nuclear matter, $\bar\psi\psi$ and $\bar\psi\gamma^{\mu}\psi$ are taken at their expectation values and the spatial components of currents vanish. The scalar and vector self-energies read: 

\begin{equation} 
S = G_S(\rho_B)\rho_s,
\end{equation}
\begin{equation} 
V_0 = G_V(\rho_B)\rho_B + \Sigma_r,
\end{equation}

\noindent where $\rho_s = \langle \bar\psi\psi\rangle $  and $\rho_B  = \langle \bar\psi\gamma^0\psi \rangle$ are the scalar and vector densities. $\Sigma_r$ is the rearrangement term in the mean field: 

\begin{equation} 
\Sigma_r = \frac{1}{2}\frac{\partial G_S}{\partial\rho_B}\rho_s^2+\frac{1}{2}\frac{\partial G_V}{\partial\rho_B}\rho_B^2. \label{rea}
\end{equation} 

\noindent It has to be noted that the rearrangement term $\Sigma_r$ is essential to ensure the density dependent models to be thermodynamically consistent. 

The self-consistent nucleon mass is related to the scalar self-energy:  

\begin{equation} 
M^*_{N} = M_N + S, \label{nucleonmass}
\end{equation} 

\noindent with $M_N$ the free nucleon mass. The spin-orbit splitting in finite nuclei are strongly related to the value of the effective nucleon mass at saturation density $\rho_{\rm sat}$. To accurately reproduce spin-orbit splitting, it should be at least reduced to the value $ M^*_N(\rho_{\rm sat})/M_N \sim 0.64$ \cite{furn}. We will consider that an acceptable range for the effective nucleon mass would be $M^*_N(\rho_{\rm sat})/M_N = 0.65 \pm 0.05$. The scalar density is given as a function of the Fermi momentum $k_F$: 

\begin{equation} 
\rho_s  =  \frac{M^*_N}{\pi^2}\left [ k_FE_F-(M^*_N)^2\ln\left(\frac{k_F+E_F}{M^*_N}\right) \right ], 
\end{equation}
 
\noindent where $E_F  = \sqrt{(k_F)^2+(M_N^*)^2}$. The nucleon density is $\rho_B = \frac{2}{3\pi^2}k_F^3$. The energy density and pressure of nuclear matter can be obtained from the energy momentum tensor

\begin{eqnarray} 
\mathcal{E} & =& \mathcal{E}_{kin} -\frac{1}{2}G_S\rho_s^2 + \frac{1}{2}G_V\rho_B^2, \label{energy}  \\
P &=& E_F\rho - \mathcal{E}_{kin} +\frac{1}{2}G_S\rho_s^2 + \frac{1}{2}G_V\rho_B^2 + \Sigma_r\rho_B,\label{pressure}
\end{eqnarray}

\noindent where the kinetic term reads: 

\begin{equation}
 \mathcal{E}_{kin} = \frac{1}{4}[ 3 E_F\rho_B + M^*_N\rho_s ].
\end{equation}

\noindent Empirically, the binding energy per nucleon, defined as $\mathcal{E}_B = \mathcal{E}/\rho_B - M_N$, should present a minimum at the saturation density $\rho_{\rm sat}$. At this density, the pressure vanishes $P(\rho_{\rm sat}) = 0$, which gives a non trivial constraint. We have considered the empirical range for saturation density $\rho_{\rm sat} = 0.153 \pm 0.015 {\rm fm^{-3}}$, with binding energy $\mathcal{E}_B(\rho_{\rm sat}) = -15.5 \pm 1.0 {\rm MeV}$. 

We will also need the nuclear matter incompressibility modulus defined as: 

\begin{equation} 
K = 9\rho_B\frac{\partial P}{\partial \rho_B}. \label{comp}
\end{equation}  

\noindent The value of $K(\rho_{\rm sat})$ is related to isoscalar giant monopole resonances in spherical nuclei. However, the empirical estimates are strongly model dependent \cite{blaizot,ma} and we will consider that a value of $K(\rho_{\rm sat}) = 250 \pm 50 {\rm MeV}$ remains acceptable.

\subsection{Determination of the density dependence of the couplings} 

The preceding point coupling model is the framework in which we want to include the quark-diquark and pionic exchange contributions to the nucleon self-energies. To achieve that, the procedure consists in mapping the self-energies in order to deduce the density dependent couplings $G_S(\rho_B)$ and $G_V(\rho_B)$. This is achieved by writing: 

\begin{eqnarray} 
G_S(\rho_B)\rho_s &=&  \Sigma_S^0 + \Sigma_S^{\pi}, \label{maps} \\ 
G_V(\rho_B)\rho_B + \Sigma_r & = & \Sigma_V^0 + \Sigma_V^{\pi},\label{mapv}
\end{eqnarray}

\noindent where $\Sigma_{S,V}^0$ are the self-energies deduced from quark-diquark picture in Section 2 and $\Sigma_{S,V}^{\pi}$ are pionic contributions derived in Section 3 ($\Sigma_r$ is defined Eq.(\ref{rea})). The self-energies coming from quark-diquark are taken in linear approximation: 

\begin{equation} 
\Sigma_S^0 \sim G_S^0\rho_s,
\end{equation} 

\begin{equation} 
\Sigma_V^0 = G_V^0\rho_B.
\end{equation}

We have verified that, using an accurate fit in powers of $k_F$ for the scalar self-energy from quark-diquark picture, does not change the results. 
The pionic contributions have been written in powers of $\rho_B$. We therefore have, considering that $\rho_s \sim \rho_B$ up to saturation density: 

\begin{eqnarray} 
G_S(\rho_B)& = & G_S^0 +\frac{1}{M_N^2}\left [c^{S}_0+ c^{S}_1\frac{\rho_B^{1/3}}{M_N} + c^{S}_2\frac{\rho_B^{2/3}}{M_N^2}\right], \label{gs}\\
G_V(\rho_B)\rho_B + \Sigma & = & \left (G_V^0 +\frac{1}{M_N^2}\left [c^{V}_0+ c^{V}_1\frac{\rho_B^{1/3}}{M_N} + c^{V}_2\frac{\rho_B^{2/3}}{M_N^2}\right]\right)\rho_B. \label{gvdiff}
\end{eqnarray} 

\noindent The equation on $G_V(\rho_B)$ is a differential equation. The solution is: 

\begin{equation} 
G_V(\rho_B) = G_V^0 + \frac{1}{M_N^2}\left[ c_0^V + \frac{6}{7}(c_1^V-\frac{1}{6}c_1^S)\frac{\rho_B^{1/3}}{M_N} + \frac{3}{4}(c_2^V-\frac{1}{3}c_2^S)\frac{\rho_B^{2/3}}{M_N^2}\right]. \label{gv}
\end{equation}

\subsection{ Parameter free results} 

We should at this point note that, as argued in \cite{bentz}, the NJL model generates a vacuum Mexican hat potential which contains an attractive tadpole diagram working against saturation. Consistently with the NJL model, we have verified that the nuclear matter cannot saturate if one takes into account only the quark-diquark part of the self-energy in the mapping procedure described above, neglecting the pionic contributions. 

We now come to the results when both quark-diquark and pion exchange contributions are taken into account in the mapping procedure. In a first step, we have applied the mapping procedure, using different couples of values $(G_S^0,G_V^0)$  corresponding to constituent quark mass values in the range $m=370 - 440 {\rm MeV}$. It has to be noted that, for a  given constituent quark mass value, there is no free parameter to fine tune on the saturation properties. The binding energy does not present any minimum for the lowest quark masses, and presents an unrealistic saturation point for $m = 440 {\rm MeV}$, with density $\rho_B \sim 0.03 {\rm fm^{-3}}$ and $\mathcal{E}_B \sim -0.5 {\rm MeV}$. While it is far from the empirical region, this result is encouraging since, with no free parameter adjusted to reproduce the saturation point, it was highly non trivial to obtain a minimum.  It is known that this minimum results from a subtle cancellation between the large scalar and vector self-energies. 

It is clear that our model suffers of some approximations, particularly the quark-diquark description in the NJL model. For example, the static approximation can lead in the vacuum to almost a $10 \%$ deviation from the exact solution of the Faddeev equation. An improvement of the quark-diquark picture would lead to a modification of the linear part of the self-energy. We explore the possibility in the next section to improve the nuclear matter description by taking into account a phenomenological correction to the quark-diquark picture of the nucleon.

\subsection{Phenomenological correction to the quark-diquark part of the self-energy} 

 In this section, we explore the possibility to improve nuclear matter saturation by adding a phenomenological correction term linear in the density to the self-energy. 

\begin{eqnarray} 
\delta\Sigma_S &=& \delta G_S^0 \rho_s, \\ 
\delta\Sigma_V & = & \delta G_V^0 \rho_B,
\end{eqnarray}

\noindent in the scalar ($S$) and vector ($V$) channels. Since we get all the contributions to the self-energy associated to pion exchange at $k_F^3$ order, this correction should be attributed to the part of the self-energy associated to the quark-diquark dynamics. In order to avoid too many free parameters, we consider the same correction for the scalar and vector self-energies, namely $\delta G_{S}^0 = \delta G^0_V = \delta G^0$. The equations (\ref{gs},\ref{gv}) are modified by adding this contribution to the couplings. 

For a given constituent quark mass value, $\delta G^0$ is adjusted, via a {\em least squares} procedure to reproduce the saturation properties which we recall below: 

\begin{itemize}
\item the saturation density $ \rho_{sat}  = 0.153 \pm 0.015 {\rm {fm}^{-3}}$
\item the binding energy $\mathcal{E}_B(\rho_{sat}) = -15.5 \pm 1 {\rm MeV}$ 
\item the nucleon effective mass $M_N^*(\rho_{sat})/M_N = 0.65 \pm 0.05 $ 
\item the nuclear matter incompressibility modulus $K(\rho_{sat}) = 250 \pm 50 {\rm MeV}$.
\end{itemize}

\begin{figure}[htb]
	{\centering
		\epsfig{file=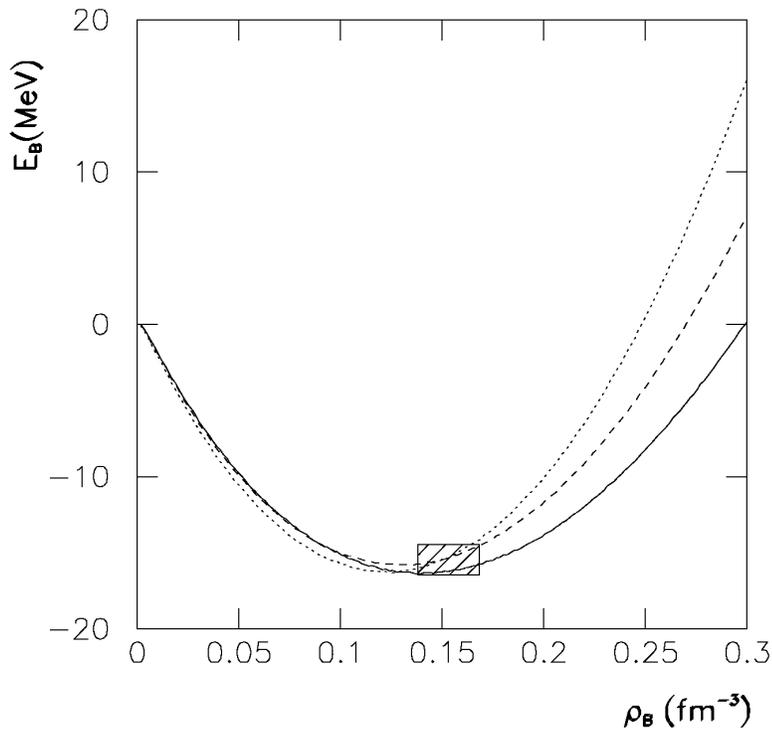,scale=0.90}
	\caption{Binding energy per nucleon versus baryonic density for three constituent quark mass values: $m=380$ MeV (solid line), $m=400$ MeV (dashed line), $m=420$ MeV (dotted line).}
	\label{ebpcm}}
\end{figure}

The results for  different quark mass values are listed in Table \ref{pcm2}. For each of these parameterizations, the average error on the four observables is less than one error bar, which is a remarkable improvement, since it has been obtained with only one free parameter. The binding energy versus baryonic density is represented on Fig. \ref{ebpcm} for three quark mass values.

\begin{table}[htb]
\begin{center}
\begin{tabular}{cccccc}
\hline\hline
$m ({\rm MeV}) $ & $\delta G^{0} (\mathrm{fm}^{2})$& $\rho_{sat} ({\rm fm^{-3}})$ & $\mathcal{E}_B ({\rm MeV})$ & $M^*_N/M_N$ & $K ({\rm MeV})$ \\
\hline
380 & -2.9 & 0.134 & -16.3 & 0.81 & 249 \\
390 & -2.8 & 0.134 & -16.2 & 0.82 & 245 \\ 
400 & -2.1 & 0.128 & -16.4 & 0.77 & 260 \\ 
410 & -1.7 & 0.122 & -16.3 & 0.75 & 267 \\ 
420 & -1.4 & 0.12 & -16.3 & 0.73 & 274 \\ 
\hline\hline
\end{tabular} 
\caption{$\delta G^0$ (see text) values and saturation properties of nuclear matter for different quark mass values.  \label{pcm2}}
\end{center} 
\end{table} 

 It is clear from Fig. \ref{ebpcm} that the binding energy is almost independent of the constituent quark mass value for densities around and below the saturation point. Although there is only one free parameter, the saturation properties are reasonable for any of the quark mass values considered. However, the saturation density is somewhat too low and the effective nucleon mass too high.

\begin{figure}[htb]
	{\centering
		\epsfig{file=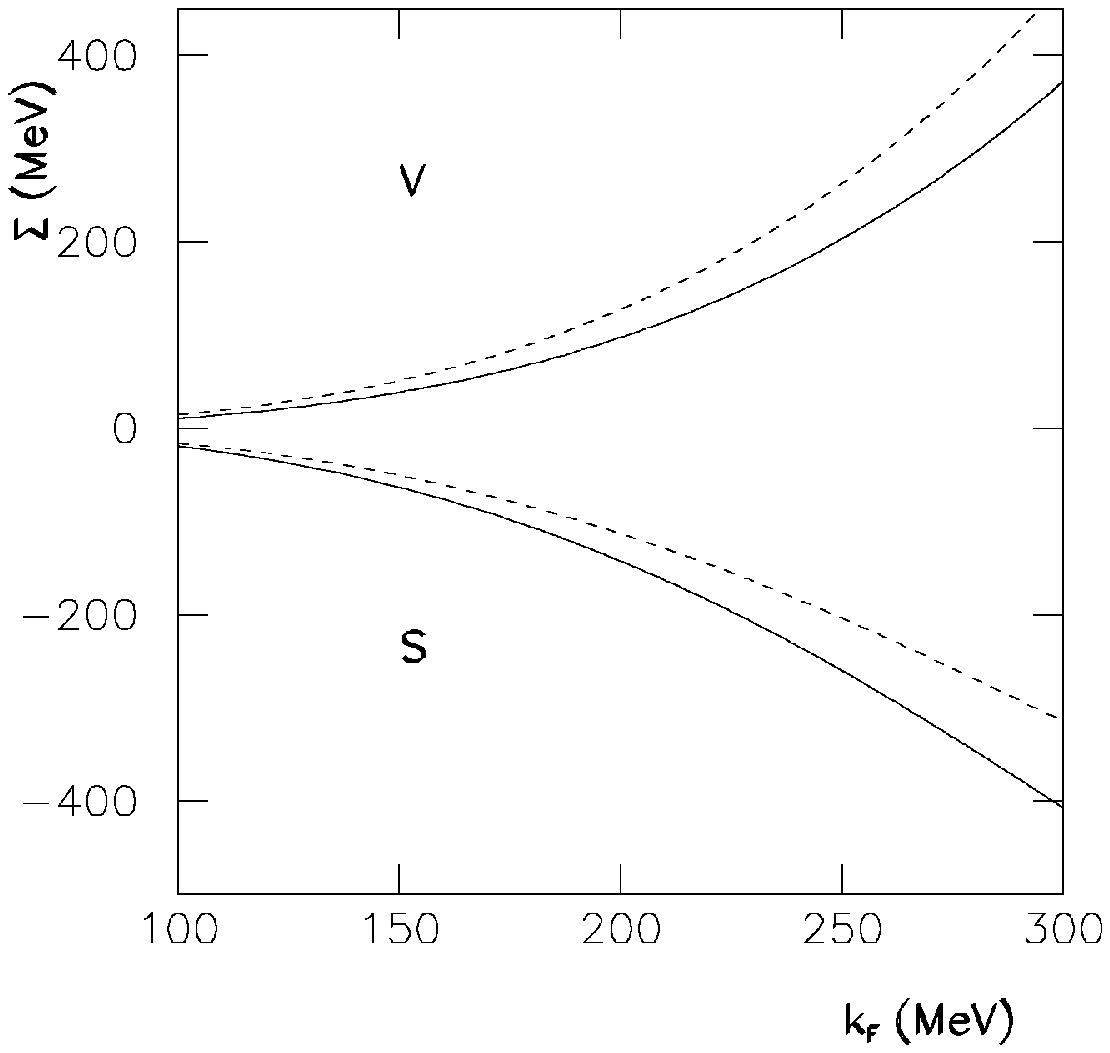,scale=0.90}
	\caption{Evolution of the self-energies as functions of $k_F$ at the different steps: quark-diquark and pionic exchange contributions alone (dashed line) and self-energy with the $k_F^3$ correction added (dotted line) for PCM1 parameterization.}
	\label{selfsdetail}}
\end{figure}

\noindent The best fit is given by the $m=410 {\rm MeV}$ parameterization with an average error of 0.7 error bar. It will be denoted below PCM1 parameterization. Figure \ref{selfsdetail} represents the evolution of the self-energies as function of $k_F$ for the PCM1 parameterization at the different steps of the calculation:  quark-diquark+ pionic exchange contributions alone and the final result. It is clear that the most important part of the self-energies is generated by the quark-diquark and pionic exchange contributions, whereas the phenomenological contribution is weaker.  Quantitatively, this can be confirmed with the ratio $\delta G^0/G_S^0$, which is of the order $0.14$.  We can conclude that it is possible to obtain realistic saturation properties with relatively small correction to the quark-diquark picture of the nucleon.

\noindent For comparison, in \cite{plohl}, the nucleon self-energies have been extracted at tree level from a realistic NN potential based on ChPT at fourth order. The authors found that most of the self-energies were generated by the contact terms. Similarly, in our approach, the quark-diquark picture is also responsible for the most important part of the self-energies. 

 Till now, the part of the nucleon self-energies from ChPT has been determined up-to and including order $k_F^5$. At three loop level of ChPT, three body interactions appear and would contribute first at order $k_F^6$. However, higher order effects coming from the quark-diquark structure of the nucleon could also contribute at $k_F^6$ order, thus the inclusion of a phenomenological $k_F^6$ contribution raises ambiguity in its interpretation. The Appendix A is dedicated to the study of the effects of the inclusion of a phenomenological $k_F^6$ term on saturation properties and a comparison of the results with a Dirac-Brueckner-Hartree-Fock calculation is reported in Appendix B.

\section{Conclusion}

We have investigated the possibility to describe infinite and symmetric nuclear matter in an approach constrained by quantum chromodynamics. We have mapped the nucleon self-energies of a point coupling relativistic mean-field model on self-energies obtained in effective theories of QCD. We have determined the contributions to in-medium nucleon self-energy by separating the short range part, driven principally by the quark structure of the nucleon, from the long range part, dictated by pion dynamics. 

We have taken the nucleon structure into account in a simple quark-diquark picture in a Nambu-Jona-Lasinio model, which is chirally invariant and reproduces the spontaneous chiral symmetry breaking. The quark-diquark picture generates large attractive scalar and repulsive vector self-energies in the medium, with an average ratio at saturation density of $\Sigma_S^0/\Sigma_V^0 \sim -2$. 

The long range part has been determined by including the one pion and iterated one pion exchange described in the framework of chiral perturbation theory at order $\mathcal{O}(k_F^5)$. The self-energies obtained are found to be approximately  $130 {\rm MeV}$ at saturation for scalar and vector channels. 

A saturation point with too low density is obtained without any free parameter to fine tune for a given constituent quark mass value. The description of nuclear matter saturation properties (saturation density, binding energy, effective nucleon mass and incompressibility modulus) is improved by introducing a correction term to the self-energy, linear in the density, which could be interpreted as a correction to the quark-diquark picture approximations. This term is found to be relatively small, and we get a reasonable description of nuclear saturation considering that, for a given quark mass value, we have only one free parameter.

To summarize, we have obtained a reasonable description of saturation properties of nuclear matter with  one free parameter.  The possibility to obtain a rather good description of saturation properties by correcting quark-diquark picture approximations with one weak phenomenological correction term linear in the density clearly indicates that, by refining the quark-diquark description of the nucleon, the model could be noticeably improved.

\begin{appendix} 

\section{Higher order chiral perturbation theory contribution} 

We examine in this appendix the influence on the saturation properties of a $k_F^6$ contribution to the self-energies. At three loop level of ChPT, three body interactions appear and would contribute first at order $k_F^6$. To our knowledge, there are not yet explicit calculations of these contributions available. We thus adopt a phenomenological point of view, introducing in the r.h.s. of  equations (\ref{gs},\ref{gvdiff}) a new term to the self-energy defined as: 

\begin{eqnarray} 
\delta \Sigma_{S}^1& =& \delta G_{S}^1 \rho_B\rho_s, \\ 
\delta \Sigma_{V}^1 & =& \delta G_V^1\rho_B^2,
\end{eqnarray} 

\noindent in the scalar $S$ and vector $V$ channels. The implicit conjecture made in the ChPT is that these terms should be sub-dominant. We indeed expect that these new contributions, adjusted to obtain the best fit of saturation properties, will be small. However, these new terms can include also higher order effects coming from the quark-diquark structure of the nucleon. 

In order to fine tune the saturation properties, the single particle potential $U \sim S+V$ is the pertinent observable to consider. The correction to $U$ is approximately driven by the sum $\delta G_S^1+\delta G_V^1$. Thus, it does not matter how the $ k_F^6$ contribution is split into scalar and vector self-energies. One can then take only one free parameter, and will obtain equivalent results by imposing different relations between $\delta G_S^1$ and $\delta G_V^1$. We thus assume that the correction to the vector self-energy is zero. 

\begin{equation} 
\delta G_V^1 = 0
\end{equation} 

\noindent We have thus only one more free parameter $\delta G_S^1$. The results are listed in Table \ref{pcm3} for different quark mass values. 

\begin{table}[htb]
\begin{center}
\begin{tabular}{ccccccc}
\hline\hline
$m ({\rm MeV})$ &  $\delta G^{0} (\mathrm{fm}^{2})$& $\delta G_S^{1} (\mathrm{fm}^{5})$&$\rho_{\rm sat} ({\rm fm^{-3}}) $ & $\mathcal{E}_B ({\rm MeV})$ & $M^*_N/M_N$ & $K ({\rm MeV})$ \\
\hline
390 & -2.5 & -5.4 & 0.168 & -15.5 & 0.76 & 231 \\ 
400 & -1.8 & -7.3 & 0.162 & -15.3 & 0.70 & 260 \\ 
410 & -1.3  & -8.5 & 0.156 & -15.4 & 0.67 & 288 \\ 
420 & -1.0  & -8.6  & 0.146 & -15.0 & 0.66 & 292 \\ 
430 & -0.8  & -7.5& 0.138 & -15.2 & 0.65 & 306 \\ 
\hline\hline
\end{tabular} 
\caption{$\delta G^0$, $\delta G_S^1$ values (see text) and saturation properties of nuclear matter for different quark mass values. \label{pcm3}}
\end{center} 
\end{table} 

\noindent The average error on saturation properties is less than $0.6$ error bar in all cases. The best fit is obtained for $m=410 {\rm MeV}$ with an average error of $0.2$ error bar on saturation properties. This is a remarkable improvement by a factor four with regard to PCM1 parameterization. The $m=410 {\rm MeV}$ set will be denoted PCM2 parameterization. 

The leading order correction term $\delta G^0$ has been reduced between PCM1 and PCM2 from $-1.7 {\rm fm^{2}}$ to $-1.3 {\rm fm^2}$. The ratio $\delta G^0/G_S^0$ is now $0.11$, and the $\delta G_S^1$ correction is relatively small with $\delta G_S^1\rho_{\rm sat}/G_S^0 \sim 0.12$. The total phenomenological correction is approximately of the same magnitude as with only a linear correction term.
\noindent Let us mention that, in \cite{plohl}, the nucleon self-energies have been extracted at tree level from a realistic nucleon-nucleon potential. In particular, the Idaho NN potential, based on ChPT up to fourth order, gives at third order a total pionic contribution of approximately the same magnitude as $\delta G_S^1$. 

\noindent We can conclude that a $k_F^6$ contribution to the self-energy, even weak, could lead to an improvement of saturation properties. In order to identify the origin of this contribution, it would be valuable to obtain an evaluation of three loop level contributions in ChPT, so as to separate effects coming from pion exchange and from quark structure of the nucleon.  

\section{Comparison with DBHF calculation from realistic NN potential} 

\begin{figure}[htb]
	{\centering
		\epsfig{file=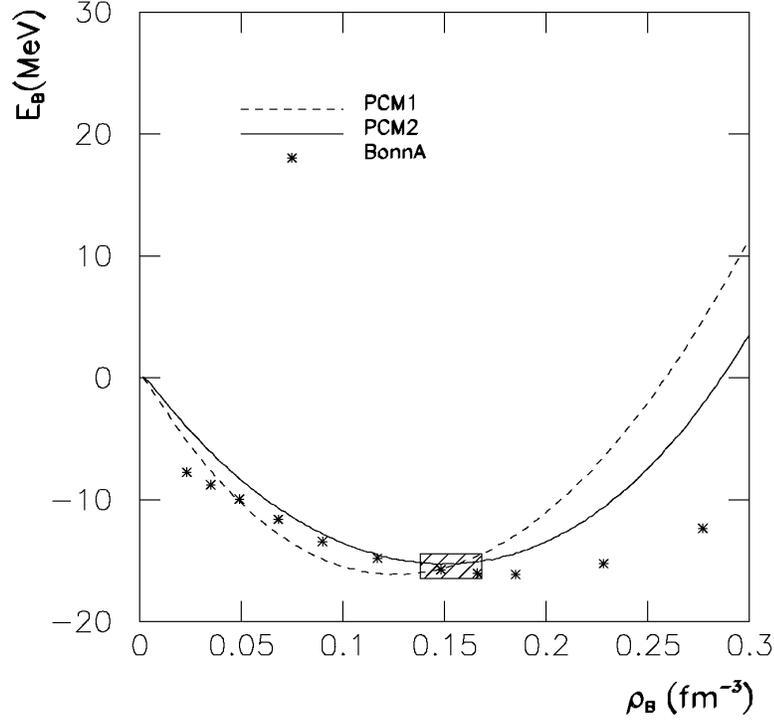,scale=0.90}
	\caption{Binding energy as functions of the baryonic density for PCM1 and PCM2 parameterizations and DBHF calculation based on BonnA potential (stars). The dashed area is the empirical region of the saturation point. }
	\label{eb}}
\end{figure}

\begin{figure}[htb]
	{\centering
		\epsfig{file=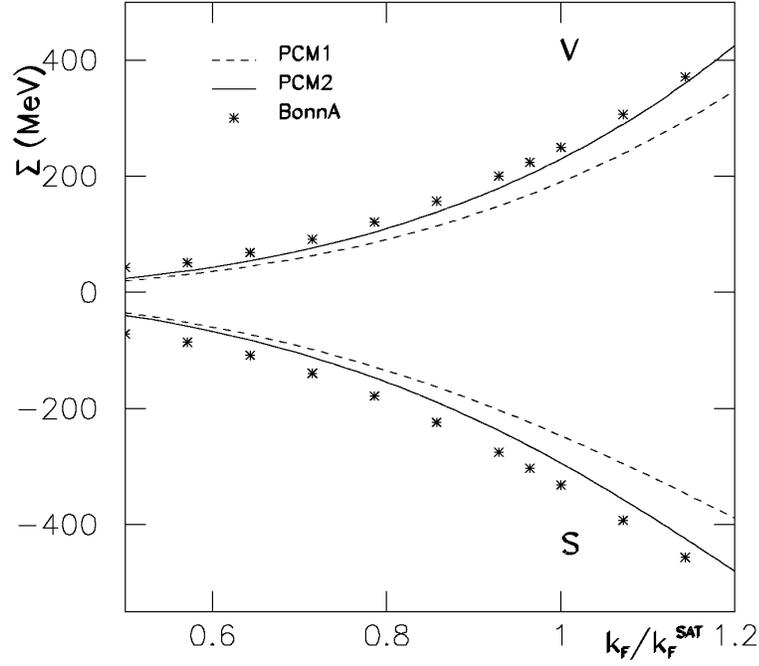,scale=0.90}
	\caption{Scalar and vector nucleon self-energies for the PCM1 and PCM2 parameterizations and from DBHF calculation based on BonnA potential, as function of the Fermi momentum $k_F$ normalised to the saturation Fermi momentum $k_F^{\rm SAT}$ of each model.}
	\label{sig}}
\end{figure}

We present here the binding energy per nucleon and the nucleon self-energies on Fig. \ref{eb} and \ref{sig} for the PCM1 and PCM2 parameterizations. The results from a Dirac-Brueckner-Hartree-Fock calculation based on BonnA potential \cite{gross} are also shown. 

\noindent The empirical saturation point region is also represented on Fig. \ref{eb} as a dashed area. It is clear that PCM1 parameterization leads to a too low saturation density, while the PCM2 saturation point is approximately in the centre of the empirical region. The DBHF calculation \cite{gross} gives a too high saturation density $\rho_{\rm sat} = 0.185 {\rm fm^{-3}}$, but it is, to our knowledge, the most elaborate DBHF calculation based on a realistic NN potential. As can be seen from Fig. \ref{eb}, the incompressibility modulus is lower for DBHF calculation ($K \sim 240 {\rm MeV}$) than for our models. 

\noindent As the three calculations give rather different saturation densities, we found more instructive to compare the self-energies plotted against the Fermi momentum, normalised to the saturation Fermi momentum of each model. This is done on Fig. \ref{sig}. As can be seen, the PCM1 parameterization gives weaker values for the self-energies, but the PCM2 parameterization leads to a good agreement with the DBHF BonnA calculation for densities in the range $\rho_B = 0.1 -1.6 \rho_{\rm sat}$.

\end{appendix}

\end{document}